\documentclass[a4paper,aps,prd,twocolumn,preprintnumbers,showpacs,superscriptaddress,nofootinbib]{revtex4}
\usepackage{graphics,graphicx,epsfig}
\usepackage{amsfonts,amsmath,amssymb}

\input{tcilatex}

\begin{document}

\title{The Brown-York mass of black holes in Warped Anti-de Sitter space}
\author{Gaston Giribet}
\email{gaston-at-df.uba.ar}
\affiliation{Physics Department, University of Buenos Aires FCEN-UBA, and IFIBA-CONICET,
Ciudad Universitaria, Pabell\'on I, 1428, Buenos Aires, Argentina.}
\author{Andr\'es Goya}
\email{af.goya-at-df.uba.ar}
\affiliation{Physics Department, University of Buenos Aires FCEN-UBA, and IFIBA-CONICET,
Ciudad Universitaria, Pabell\'on I, 1428, Buenos Aires, Argentina.}

\begin{abstract}
We give a direct computation of the mass of black holes in Warped Anti-de
Sitter space (WAdS) in terms of the Brown-York stress-tensor at the
boundary. This permits to explore to what extent the holographic
renormalization techniques can be applied to such type of deformation of
AdS. We show that, despite some components of the boundary stress-tensor
diverge and resist to be regularized by the introduction of local
counterterms, the precise combination that gives the quasilocal energy
density yields a finite integral. The result turns out to be in agreement
with previous computations of the black hole mass obtained with different
approaches. This is seen to happen both in the case of Topologically Massive
Gravity and of the so-called New Massive Gravity. Here, we focus our
attention on the latter. We observe that, despite other conserved charges
diverge in the near boundary limit, the finite part in the large radius
expansion captures the physically relevant contribution. We compute the
black hole angular momentum in this way and we obtain a result that is in
perfect agreement with previous calculations.
\end{abstract}

\maketitle

\section{Introduction}

The idea of extending AdS/CFT correspondence to Warped AdS spaces (WAdS) has
been originally proposed in Ref. \cite{Strominger}, and it was further
studied in Refs. \cite{Compere, Compere2, Henneaux, Guica, AnninoEsoleGuica,
Anninos, Belgrado, Belgrado2, Olea, Hofman}. This represents one of the most
appealing attempts to generalize holography to non-AdS spaces, and this is
because WAdS spaces appear in several contexts. For instance, WAdS spaces
provide gravity duals for condensed matter systems with Schr\"{o}dinger
symmetry \cite{McGreevy, Son}, they are closely related to the geometry of
rotating black holes \cite{Sandin, KerrCFT}, and they also appear in
relation to many other interesting subjects \cite{Godel, Song, inner}.
Asymptotically WAdS$_{3}$ spaces turn out to be exact solutions of String
Theory \cite{Strings, Strings4, Strings5, Strings3, Strings2} as well as of
other models of three-dimensional gravity, including Higher-Spin Gravity 
\cite{Grumiller}, Topologically Massive Gravity (TMG) \cite{Clement,
Clement2}, and New Massive Gravity (NMG) \cite{Clement3}. Here, we will be
concerned with the latter: We will consider asymptotically WAdS$_{3}$ black
holes in NMG. For such solutions, we will give a direct computation of the
mass in terms of the Brown-York stress-tensor \cite{BrownYork} in the
boundary of the space. We do this to explore to what extent the holographic
renormalization techniques can be applied to such a deformation of AdS.
Whether or not the Brown-York tensor can be defined at the boundary of WAdS$%
_{3}$ space is a question that has been raised, for instance, in Ref. \cite%
{Guica}. Here, we will show that, despite some components of the Brown-York
stress-tensor diverge in the near boundary limit and resist to be
regularized by the introduction of local counterterms, the integral of the
precise combination that gives the definition of the quasilocal energy as a
conserved charge yields a finite integral. The result we obtain happens to
be in agreement with computations of the black hole mass obtained by
different methods \cite{Clement, Tonni, Coreanos}. Finiteness of the
conserved charge computed in this way follows from cancellations that occur
near the boundary. In contrast to the mass, in the case of the angular
momentum the charge associated to it can not be regularized by the
introduction of local boundary counterterms. However, the finite part in the
near boundary expansion happens to capture the physically relevant
information, and it is shown to exactly reproduce the black hole angular
momentum.

The paper is organized as follows: In Section II, we briefly review the
theory of New Massive Gravity introduced in Ref. \cite{NMG}. In Section III,
we discuss the geometry of Warped Anti-de Sitter space and black holes that
asymptote to it. In Section IV, we study boundary terms and the Brown-York
stress-tensor they induce. We consider the near boundary limit of this
stress-tensor and use it in Section V to calculate the mass of the Warped
Anti-de Sitter black holes. That is, we compute the Brown-York quasilocal
energy in the limit that the boundary tends to spatial infinity. The result
we obtain is in agreement with previous computations. We also discuss the
analogous computation in the case of the gravitational Chern-Simons term
being added.

\section{New massive gravity}

Let us begin by reviewing New Massive Gravity theory \cite{NMG}. The action
of the theory consists of the sum of three different contributions, namely%
\begin{equation}
S=S_{\text{EH}}+S_{\text{NMG}}+S_{\text{B}},  \label{action}
\end{equation}%
where the first term is the Einstein-Hilbert action with cosmological
constant, 
\begin{equation}
S_{\text{EH}}=\frac{1}{16\pi G}\int_{\Sigma }d^{3}x\sqrt{-g}\left(
R-2\Lambda \right) ,
\end{equation}%
and the second term contains contributions of higher order,%
\begin{eqnarray}
S_{\text{NMG}} &=&\frac{1}{16\pi G}\int_{\Sigma }{d^{3}x\sqrt{-g}\left(
f^{\mu \nu }G_{\mu \nu }-\right. }  \notag \\
&&{\frac{1}{4}m^{2}(f_{\mu \nu }f^{\mu \nu }-f^{2}))},  \label{SNMG}
\end{eqnarray}%
where $G_{\mu \nu }\ $is the Einstein tensor $G_{\mu \nu }=R_{\mu \nu }-%
\frac{1}{2}Rg_{\mu \nu }$, and field $f_{\mu \nu }$ is a rank-two auxiliary
field which, after being integrated, gives%
\begin{equation}
f_{\mu \nu }=\dfrac{2}{m^{2}}\left( R_{\mu \nu }-\dfrac{1}{4}Rg_{\mu \nu
}\right) .  \label{f}
\end{equation}

The third term in (\ref{action}), S$_{B}$, is a boundary action needed for
the variational principle to be defined in a specific way. We will discuss
the boundary terms later.

By reinserting (\ref{f}) back in (\ref{SNMG}) the higher-curvature terms
take the form 
\begin{equation}
S_{\text{NMG}}=\dfrac{1}{16\pi Gm^{2}}\int_{\Sigma }{d^{3}x\sqrt{-g}}\left( {%
R_{\mu \nu }R^{\mu \nu }-\dfrac{3}{8}R^{2}}\right) ,  \label{B}
\end{equation}%
which is the form of the action presented in \cite{NMG}.

The equations of motion derived from action (\ref{action}) read 
\begin{equation}
16\pi G\frac{\delta S}{\delta g^{\mu \nu }}=G_{\mu \nu }+\Lambda g_{\mu \nu
}+\frac{1}{2m^{2}}K_{\mu \nu }=0.  \label{eom}
\end{equation}%
which, apart from the Einstein tensor $G_{\mu \nu }$, involve the tensor%
\begin{eqnarray}
K_{\mu \nu } &=&2\square {R}_{\mu \nu }-\frac{1}{2}\nabla _{\mu }\nabla
_{\nu }{R}-\frac{1}{2}\square {R}g_{\mu \nu }+4R_{\mu \alpha \nu \beta
}R^{\alpha \beta }-  \notag \\
&&\frac{3}{2}RR_{\mu \nu }-R_{\alpha \beta }R^{\alpha \beta }g_{\mu \nu }+%
\frac{3}{8}R^{2}g_{\mu \nu }.
\end{eqnarray}

The precise combination of the square-curvature terms in (\ref{B}), $g^{\mu
\nu }K_{\mu \nu }=R_{\mu \nu }R^{\mu \nu }-(3/8)R^{2},$ is such that the
trace of the equations of motion (\ref{eom}) does not involve the mode $%
\square {R}$. This is one of the reasons why NMG is free of ghosts -- for
instance -- about flat space.

Equations of motion (\ref{eom}) are solved by all solutions of General
Relativity, provided an adequate renormalization of the effective
cosmological constant. The theory also admits solutions that are not
Einstein spaces; these have $K_{\mu \nu }\neq 0$. Probably the simplest
solutions of this sort are WAdS$_{3}$ spaces.

\section{Warped Anti-de Sitter}

\subsection{WAdS$_{3}$ space}

WAdS$_{3}$ spaces are squashed or stretched deformations of AdS$_{3}$ \cite%
{Sandin}. Such a deformation is obtained by first writing AdS$_{3}$ as a
Hopf fibration of $\mathbb{R}$ over AdS$_{2}$ and then multiplying the fiber
by constant warp factor $K$. More precisely, one first considers the metric
of AdS$_{3}$ written in coordinates%
\begin{equation}
ds^{2}=\frac{l^{2}}{4}\left( -\cosh ^{2}x\ d\tau ^{2}+dx^{2}+(dy+\sinh x\
d\tau )^{2}\right)  \label{AdS3}
\end{equation}%
and then deforms it as follows 
\begin{equation}
ds^{2}=\frac{l_{K}^{2}}{4}\left( -\cosh ^{2}x\ d\tau ^{2}+dx^{2}+K(dy+\sinh
x\ d\tau )^{2}\right) ,  \label{WAdS3}
\end{equation}%
where $x,y,\tau \in \mathbb{R}$, and $K\in \mathbb{R}$. It is usual to
parameterize the deformation by a positive constant $\nu $ defined by $%
K=4\nu ^{2}/(\nu ^{2}+3)$, such that $\nu =1$ corresponds to undeformed
--unwarped-- AdS$_{3}$. Through the deformation, the AdS$_{3}$ radius $l$
gets also rescaled as $l^{2}\rightarrow l_{K}^{2}=4l^{2}/(\nu ^{2}+3)$.
Spaces (\ref{WAdS3}) with $\nu ^{2}>1$ describe stretched AdS$_{3}$ spaces,
while those with $\nu ^{2}<1$ describe squashed deformations of it. Through
a double Wick rotation $x,\tau \rightarrow ix,i\tau $ one goes from the
spacelike WAdS$_{3}$ metric (\ref{WAdS3}) to a timelike analog of it. Here
we will be involved with spacelike stretched WAdS$_{3}$ spaces.

\subsection{WAdS$_{3}$/CFT$_{2}$}

The warping deformation breaks the $SL(2,\mathbb{R})\times SL(2,\mathbb{R})$
isometry group of AdS$_{3}$ space down to $SL(2,\mathbb{R})\times U(1)$. As
a consequence, also the asymptotic isometry group, which in the case of AdS$%
_{3}$ coincides with the two-dimensional local conformal group, gets
altered. It has been recently understood that the asymptotic group of WAdS$%
_{3}$ turns out to be generated by the semi-direct product of one copy of
Virasoro algebra and an affine extension of $u(1)$ algebra with
non-vanishing central extension; see \cite{Compere, Compere2, Belgrado,
Belgrado2, Henneaux}. That is, the asymptotic isometry group in WAdS$_{3}$
spaces certainly differs from the two-dimensional conformal group;
nevertheless, it has been shown in \cite{Hofman} that, under certain
circumstances, the symmetry results powerful enough to constrain the dual
theory and extract relevant information from it. The holographic description
of WAdS$_{3}$ black hole thermodynamics carried out in \cite{Hofman} is a
notable realization of this idea.

Motivated by the similarities between asymptotically WAdS$_{3}$ and
asymptotically AdS$_{3}$ spaces, the authors of \cite{Strominger} proposed
the idea of extending AdS/CFT to the former case. The conjecture is that
quantum gravity in asymptotically WAdS$_{3}$ space would be dual to a
two-dimensional theory which exhibits partial conformal symmetry, it being
symmetric under right -- but not left -- dilations. The main motivation we
have to study the holographic renormalization techniques in this context
comes from trying to determine to what extent what we know about holography
can be applied with no major modification to WAdS spaces as well.

\subsection{WAdS$_{3}$ black holes}

One of the most attractive properties of WAdS space is that it admits black
holes that asymptote to it and, on the other hand, are given by discrete
quotients of WAdS$_{3}$ itself. This is analogous to what happens with the Ba%
\~{n}ados-Teiltelboim-Zanelli (BTZ) black hole \cite{BTZ}, which is locally
equivalent to AdS$_{3}$. The existence of WAdS$_{3}$ black holes is very
interesting since, if thought of within the context of a WAdS$_{3}$/CFT$_{2}$
correspondence, it gives raise the hope to investigate black hole physics in
a totally new setup.

The metric of WAdS$_{3}$ black holes is given by%
\begin{eqnarray}
ds^{2} &=&dt^{2}+\left( 2\nu r-\sqrt{(\nu ^{2}+3)r_{+}r_{-}}\right)
dtd\varphi +  \notag \\
&&l^{2}\left( (\nu ^{2}+3)(r-r_{+})(r-r_{-})\right) ^{-1}dr^{2}+  \notag \\
&&\dfrac{r}{4}\left( 3(\nu ^{2}-1)r+(\nu ^{2}+3)(r_{+}+r_{-})\right. - 
\notag \\
&&4\nu \sqrt{r_{+}r_{-}(\nu ^{2}+3)})d\varphi ^{2},  \label{there}
\end{eqnarray}%
where $t\in \mathbb{R}$, the angular coordinate $\varphi \in \lbrack 0,2\pi
) $, it being identified as $\varphi \sim \varphi +2\pi $, and $r\in \mathbb{%
R}_{\geq 0}$. $r_{+}$ and $r_{-}$ are two integration constants that, for $%
r_{+}\geq r_{-}\geq 0$, represent the location of the outer and inner
horizons of the black hole. Solutions (\ref{there}) asymptote spacelike
stretched WAdS$_{3}$ at large $r$. Metric (\ref{there}) can also be written
in the ADM like form%
\begin{equation}
ds^{2}=-N_{t}^{2}dt^{2}+\rho ^{2}\left( d\varphi +N^{\varphi }dt\right) ^{2}+%
\dfrac{l^{2}dr^{2}}{4\rho ^{2}N_{t}^{2}},  \label{FF}
\end{equation}%
with 
\begin{eqnarray}
\rho ^{2} &=&\dfrac{r}{4}(3(\nu ^{2}-1)r+(\nu ^{2}+3)(r_{+}+r_{-})-  \notag
\\
&&4\nu \sqrt{r_{+}r_{-}(\nu ^{2}+3)}),  \label{ro} \\
N_{t}^{2} &=&\dfrac{(\nu ^{2}+3)(r-r_{+})(r-r_{-})}{4\rho ^{2}}, \\
N^{\varphi } &=&\dfrac{2\nu r-\sqrt{(\nu ^{2}+3)r_{+}r_{-}}}{2\rho ^{2}}.
\end{eqnarray}

As mentioned, WAdS$_{3}$ black holes are specific identifications of the WAdS%
$_{3}$ space \cite{Strominger}. That is, black hole geometry (\ref{there})
is constructed as a quotient of WAdS$_{3}$ space by a discrete subgroup of $%
SL(2,\mathbb{R})\times U(1)$, identifying points of the original manifold
along a direction $\partial _{\varphi }=\pi l(J_{2}/\beta _{\text{L}}-\bar{J}%
_{2}/\beta _{\text{R}})$, with $J_{2}\in SL(2,\mathbb{R})$ and $\bar{J}%
_{2}\in U(1)$, and $\beta _{\text{L},\text{R}}\in \mathbb{R}$. This allows
to define the left- and right-temperature as the inverse of the periods $%
\beta _{\text{R},\text{L}}$; namely 
\begin{eqnarray*}
T_{\text{L}} &=&\beta _{\text{L}}^{-1}=\frac{(\nu ^{2}+3)}{8\pi l^{2}}%
(r_{+}+r_{-}-\frac{1}{\nu }\sqrt{(\nu ^{2}+3)r_{+}r_{-}}), \\
T_{\text{R}} &=&\beta _{\text{R}}^{-1}=\frac{(\nu ^{2}+3)}{8\pi l^{2}}%
(r_{+}-r_{-}).
\end{eqnarray*}%
\qquad \qquad

Because of being locally equivalent to WAdS$_{3}$ space (\ref{WAdS3}), and
despite having a richer causal structure, the local geometry of black holes (%
\ref{there}) is remarkably simple. In particular, the curvature scalars
result to be independent of the integration constants $r_{\pm }$. Moreover,
all the curvature invariants turn out to be constant, given only in terms of
parameters $\nu $ and $l$; for instance,%
\begin{eqnarray*}
R=-\frac{6}{l^{2}},\quad &&R_{\mu \nu }R^{\mu \nu }=\frac{6}{l^{4}}(3-2\nu
^{2}+\nu ^{4}), \\
R_{\mu \nu }R_{\rho }^{\nu }R^{\rho \mu } &=&-\frac{6}{l^{6}}(9-9\nu
^{2}+3\nu ^{4}+\nu ^{6}).
\end{eqnarray*}

As we will see, this geometric simplicity of WAdS$_{3}$ black holes is,
paradoxically, one of the aspects that make difficult to deal with them.

\subsection{NMG\ WAdS$_{3}$ black holes}

It has been shown in \cite{Clement3} that WAdS$_{3}$ black holes (\ref{there}%
) solve the equations of motion of NMG if the parameters satisfy the
relations 
\begin{equation}
m^{2}=-\dfrac{(20\nu ^{2}-3)}{2l^{2}},\ \ \Lambda =-\dfrac{m^{2}(9-48\nu
^{2}+4\nu ^{2})}{(9-120\nu ^{2}+400\nu ^{4})}.  \label{SSS}
\end{equation}

The same type of solution for the case of NMG theory coupled to TMG was
studied in Ref. \cite{Tonni}.

Entropy of WAdS$_{3}$ black hole in NMG can be evaluated by means of Wald
formula \cite{Wald}, yielding%
\begin{equation}
S=\frac{8\pi \nu ^{3}}{(20\nu ^{2}-3)G}(r_{+}-\frac{1}{2\nu }\sqrt{(\nu
^{2}+3)r_{+}r_{-}}).  \label{SS}
\end{equation}

Remarkably, the entropy results proportional to $T_{\text{L}}+T_{\text{R}}$,
which means that it admits to be written in the Cardy like form%
\begin{equation}
S=\frac{\pi ^{2}l}{3}c(T_{\text{L}}+T_{\text{R}})
\end{equation}%
with $c$ being independent of $r_{\pm }$. Then, one may identify the central
charge of the dual theory to be%
\begin{equation}
c=\frac{96\nu ^{3}l}{(20\nu ^{4}+57\nu ^{2}-9)G}.  \label{c}
\end{equation}

Notice that, in the limit $\nu \rightarrow 1$, central charge (\ref{c})
tends to its AdS$_{3}$ value $c=24l/(17G)$; recall that in NMG the
Brown-Henneaux central charge $3l/(2G)$ gets multiplied by a factor $%
1+1/(2m^{2}l^{2})$, and, according (\ref{SSS}), $\nu =1$ corresponds to $%
m^{2}l^{2}=-17/2$.

\section{Brown-York stress-tensor}

\subsection{ADM decomposition}

Now, let us analyze the definition of the Brown-York tensor in NMG. This has
been originally studied in Ref. \cite{HohmTonni}. To define the
stress-tensor, it is convenient to write the metric in its ADM decomposition
for the radial coordinate, $r$, namely 
\begin{equation}
ds^{2}=N^{2}dr^{2}+\gamma _{ij}(dx^{i}+N^{i}dr)(dx^{j}+N^{j}dr),  \label{ADM}
\end{equation}%
where $N^{2}$ is the radial lapse function, and $\gamma _{ij}$ is the
two-dimensional metric on the constant-$r$ surfaces. The Latin indices $%
i,j=0,1$, refer to the coordinates on the constant-$r$ surfaces, while the
Greek indices are $\mu ,\nu =0,1,2$, and include the radial direction $r$ as
well. In the case of asymptotically AdS$_{3}$ spaces, one knows how to
restrict the $r$-dependence of $\gamma _{ij}$ as it comes from the
Fefferman-Graham expansion \cite{FG}, which in three-dimensions results
consistent with the Brown-Henneaux asymptotic boundary conditions \cite{BH}.
For WAdS$_{3}$ the asymptotic boundary conditions were studied in Refs. \cite%
{Compere, Compere2, Belgrado, Belgrado2, Henneaux} for the case of TMG; in
particular, it has been shown in \cite{Henneaux} that the theory admits more
than one set of consistent boundary conditions, all of them being defined in
a way that WAdS$_{3}$ black hole solutions (\ref{there}) are gathered. We
assume such kind of asymptotic behavior. More precisely, we consider
perturbation of the $r_{+}=r_{-}=0$ configuration (\ref{there}) of the form 
\begin{eqnarray}
ds^{2} &=&dt^{2}+2\nu rdtd\varphi +\frac{l^{2}dr^{2}}{r^{2}(\nu ^{2}+3)}+ 
\notag \\
&&\dfrac{3r}{4}(\nu ^{2}-1)d\varphi ^{2}+h_{\mu \nu }dx^{\mu }dx^{\nu },
\end{eqnarray}%
gathering metrics with falling-off conditions%
\begin{eqnarray*}
h_{rr} &\simeq &\mathcal{O}(r^{-3}),\quad h_{\varphi \varphi }\simeq 
\mathcal{O}(r), \\
\quad h_{t\varphi } &\simeq &\mathcal{O}(1),\quad h_{tt}\simeq \mathcal{O}%
(r^{-3}).
\end{eqnarray*}

\subsection{Boundary terms}

Boundary terms $S_{\text{B}}$ are introduced in (\ref{action}) for the
variational principle to be defined in such a way that both the metric $%
g_{\mu \nu }$ and the auxiliary field $f_{\mu \nu }$ are fixed on the
boundary $\partial \Sigma $. With this prescription, the boundary action $S_{%
\text{B}}$ reads%
\begin{equation}
S_{\text{B}}=-\frac{1}{8\pi G}\int_{\partial \Sigma }d^{2}x\sqrt{-\gamma }%
\left( K+\frac{1}{2}\hat{f}^{ij}(K_{ij}-\gamma _{ij}K)\right) .  \label{Sb}
\end{equation}%
Here, $\gamma _{ij}$ is the metric induced on $\partial \Sigma $ and $K_{ij}$
is the extrinsic curvature, with $K=\gamma ^{ij}K_{ij}$. On the other hand, $%
\hat{f}^{ij}$ in (\ref{Sb})\ comes from decomposing the contravariant field $%
f^{\mu \nu }$ as%
\begin{equation*}
f^{\mu \nu }=\left( 
\begin{array}{cc}
f^{ij} & h^{j} \\ 
h^{i} & s%
\end{array}%
\right)
\end{equation*}
and defining 
\begin{equation*}
\hat{f}^{ij}\equiv f^{ij}+2h^{(i}N^{j)}+sN^{i}N^{j},\quad \hat{f}\equiv
\gamma _{ij}\hat{f}^{ij}.
\end{equation*}

The first term in (\ref{Sb}) corresponds to the Gibbons-Hawking term. The
other two terms come from the higher-curvature terms of NMG. Notice that in (%
\ref{Sb}) the field $\hat{f}^{ij}$ couples to the Israel tensor $%
K_{ij}-\gamma _{ij}K$ in the same manner as the field $f^{\mu \nu }$ couples
to the Einstein tensor in the bulk action (\ref{SNMG}).

Then, the Brown-York stress-tensor can be obtained by varying action (\ref%
{action}) with respect to the metric $\gamma ^{ij}$; namely%
\begin{equation}
T_{ij}=\frac{2}{\sqrt{-\gamma }}\frac{\delta S}{\delta \gamma ^{ij}}_{|r=%
\text{const}}.  \label{Tij}
\end{equation}

This yields two distinct contributions, $T^{ij}=T_{\text{EH}}^{ij}+T_{\text{%
NMG}}^{ij}$. First, we have the Israel term%
\begin{equation*}
T_{\text{EH}}^{ij}=\frac{1}{8\pi G}(K^{ij}-K\gamma ^{ij}),
\end{equation*}%
and, secondly, we have the contribution coming from the higher-curvature
terms \cite{HohmTonni}%
\begin{eqnarray*}
T_{\text{NMG}}^{ij} &=&-\frac{1}{8\pi G}\left( \frac{1}{2}\hat{f}%
K^{ij}+\nabla ^{(i}\hat{h}^{j)}-\frac{1}{2}\nabla _{r}\hat{f}^{ij}+K_{k}^{(i}%
\hat{f}^{j)k}\right. - \\
&&\left. \frac{1}{2}\hat{s}K^{ij}-\gamma ^{ij}(\nabla _{k}\hat{h}^{k}-\frac{1%
}{2}\hat{s}K+\frac{1}{2}\hat{f}K-\frac{1}{2}\nabla _{r}\hat{f})\right) ,
\end{eqnarray*}%
where $\hat{h^{i}}=N(h^{i}+sN^{i}N^{j})$, $\hat{s}=N^{2}s$, and where the
covariant $r$-derivative $\nabla _{r}$ is defined as follows 
\begin{eqnarray*}
\nabla _{r}\hat{f}^{ij} &=&\frac{1}{N}\left( \partial _{r}\hat{f}%
^{ij}-N^{k}\partial _{k}\hat{f}^{ij}+2\hat{f}^{k(i}\partial
_{k}N^{j)}\right) , \\
\nabla _{r}\hat{f} &=&\frac{1}{N}\left( \partial _{r}\hat{f}-N^{k}\partial
_{k}\hat{f}\right) .
\end{eqnarray*}

\subsection{Counterterms}

The next step towards the definition of the boundary stress-tensor is adding
counterterms to regularize (\ref{Tij}) in the limit $r\rightarrow \infty $.
In asymptotically AdS$_{3}$ space this is achieved by the holographic
renormalization recipe, which amounts to add boundary terms that only
involve intrinsic boundary quantities. Here, such terms would be of the form 
\begin{equation}
S_{\text{C}}=\frac{1}{8\pi G}\int_{\partial \Sigma }d^{2}x\sqrt{-\gamma }%
(a_{0}+a_{1}\ \hat{f}+a_{2}\ \hat{f}^{2}+b_{2}\ \hat{f}_{ij}\hat{f}^{ij}+...)
\label{Sc}
\end{equation}%
The ellipses stand for higher-order intrinsic terms. From the boundary
viewpoint these terms are thought of as counterterms in the dual theory;
meaning that the renormalized boundary stress-tensor is defined by taking
the $r\rightarrow \infty $ limit of the improved stress-tensor%
\begin{equation}
T_{ij}\rightarrow T_{ij}^{\ast }=T_{ij}+\frac{2}{\sqrt{-\gamma }}\frac{%
\delta S_{\text{C}}}{\delta \gamma ^{ij}}.  \label{Tijren}
\end{equation}

The choice of counterterms (\ref{Sc}), namely the choice of coefficients $%
a_{i},b_{i}$, is partially determined by demanding the action to be finite.
Regarding this point, it is worthwhile mentioning a peculiarity of WAdS$_{3}$
space, which is the fact that WAdS$_{3}$ space does not admit a real
Euclidean section. Therefore, one has to specify precisely what does
requiring finiteness in the action actually mean in this context. We will
circumvent the problem by saying that here we are dealing with stationary
solutions, so we will demand the Lorentzian action integrated over a finite
time interval to be finite. This is achieved by choosing%
\begin{equation}
a_{0}=-\dfrac{8\nu ^{2}\sqrt{\nu ^{2}+3}}{(20\nu ^{2}-3)l}.  \label{Con}
\end{equation}

Nevertheless, one may ask whether (\ref{Con}) is the only possible choice.
It turns out that the answer is in the affirmative: In contrast to what
happens with other solutions of massive gravity, like the ones found in
Refs. \cite{Hairy} and \cite{Lifshitz}, whose boundary stress-tensors can be
regularized by introducing additional counterterms, here the geometrical
simplicity of the WAdS$_{3}$ spaces happens to play against us: For WAdS$%
_{3} $ we have%
\begin{eqnarray*}
\hat{f}=-\frac{2\nu ^{2}}{m^{2}l^{2}},\ \ \ &&\hat{f}_{ij}\hat{f}^{ij}=\frac{%
2}{m^{4}l^{4}}(9-18\nu ^{2}+10\nu ^{4}), \\
\hat{f}_{ij}\hat{f}_{k}^{j}\hat{f}^{ki} &=&-\frac{2\nu ^{2}}{m^{6}l^{6}}%
(27-54\nu ^{2}+28\nu ^{4}),\ ...
\end{eqnarray*}%
and, therefore, there are no many options as all the terms are constant.
Thus, the frugal menu of counterterms we have at hand has only one
independent option, say (\ref{Con}). This is precisely what we meant when we
said that the geometric simplicity of WAdS$_{3}$ black holes is one of the
aspects that make difficult to deal with them.

Notice that counterterm (\ref{Con}) is consistent with the fact that WAdS$%
_{3}$ black holes reduce to BTZ\ black hole (in a rotating frame) when $\nu
=1$. For $\nu =1$ we have $a_{0}=-16/(17l)$, which is the expected value for
the case in which there is no warping and WAdS$_{3}$ space reduces to AdS$%
_{3}$. Recall that $m^{2}=-(20\nu ^{2}-3)/(2l^{2})$, so that for $\nu =1$ we
have $2l^{2}m^{2}=-17$; on the other hand, in NMG\ the counterterm needed to
regularize the boundary stress-tensor in AdS$_{3}$ is a boundary
cosmological term with coefficient $%
a_{0}=-(1+2m^{2}l^{2})/(2m^{2}l^{3})=-16/(17l)$.

\section{Conserved charges}

\subsection{Quasilocal energy}

Once the stress-tensor has been improved by adding to it the boundary
contributions $S_{\text{C}}$ that render the Lagrangian finite, one can
define the conserved charges as follows \cite{BrownYork}%
\begin{equation}
Q[\xi ]=\int ds\ u^{i}T_{ij}^{\ast }\xi ^{j},  \label{carga}
\end{equation}%
where $ds$ is the line element of the constant-$t$ surfaces at the boundary, 
$u$ is a unit vector orthogonal to the constant-$t$ surfaces, and $\xi $ is
the Killing vector that generates the isometry in $\partial \Sigma $ to
which the charge is associated. In the case of the mass, the components of
this vector are $\xi ^{i}=N_{t}u^{i}$, where the lapse function $N_{(r)}^{t}$
in (\ref{FF}). This defines the energy density;\ see \cite{BrownYork,
BalasubramanianKraus} for discussions. From (\ref{FF}) we see that the line
element $ds$ in the case we are interested in is simply $ds=\rho d\varphi $.

However, before going further, let us express a concern about the finiteness
of (\ref{carga}). This is because the fact that counterterm (\ref{Sc})
achieves to make the action finite in the way we discussed it, does not
necessarily imply that the stress-tensor is finite as well. In fact, it can
be explicitly verified that the inclusion of counterterm (\ref{Sc}) with (%
\ref{Con}) in the case of WAdS$_{3}$ black holes does not suffice to make
all the components of $T_{ij}^{\ast }$ finite. Nevertheless, it turns out
that, despite the divergences in the improved stress-tensor, the charge (\ref%
{carga}) defined with $\xi =N_{t}u$ at the boundary $r=\infty $ results
finite. It gives%
\begin{equation}
M=\dfrac{\nu ^{2}(\nu ^{2}+3)}{2(20\nu ^{2}-3)lG}\left( r_{+}+r_{-}-\dfrac{1%
}{\nu }\sqrt{(\nu ^{2}+3)r_{+}r_{-}}\right) .  \label{Mass}
\end{equation}

This agrees with the result obtained in \cite{Clement3, Tonni, Coreanos} up
to a factor $1/2$. The comparison with the mass computed in \cite{Clement3}
is discussed in Appendix C of Ref. \cite{Tonni}. Finiteness of (\ref{Mass})
follows from cancellations that take place in the near boundary limit. This
can be verified by the large $r$ expansion of the $T_{ij}^{\ast }$
components 
\begin{eqnarray*}
T_{tt}^{\ast } &\simeq &t_{tt}^{(0)}+t_{tt}^{(-1)}r^{-1}+t_{tt}^{(-2)}r^{-2}+%
\mathcal{O}(r^{-3}), \\
T_{t\varphi }^{\ast } &\simeq &t_{t\varphi }^{(1)}r+t_{t\varphi
}^{(0)}+t_{t\varphi }^{(-1)}r^{-1}+\mathcal{O}(r^{-2}),
\end{eqnarray*}%
where $t_{ij}^{(n)}$ are constant coefficients that can be found in the
Appendix, and the large $r$ expansion of the unit vector components%
\begin{eqnarray*}
u^{t} &\simeq &u_{(0)}^{t}+u_{(-1)}^{t}r^{-1}+\mathcal{O}(r^{-2}), \\
u^{\varphi } &\simeq &u_{(-1)}^{\varphi }r^{-1}+\mathcal{O}(r^{-2}).
\end{eqnarray*}%
Coefficient $u_{(-1)}^{t}$ results proportional to (\ref{Mass}).

As mentioned in Ref. \cite{GiribetLeston} in a similar context, mass formula
(\ref{Mass}) is cumbersome enough for not to doubt about its calculation by
means of (\ref{Tijren})-(\ref{carga}) actually makes sense. Nevertheless, to
convince ourselves about it, let us revise the same type of calculation for
TMG and see that it also works when the gravitational Chern-Simons term is
included.

\subsection{Gravitational Chern-Simons term}

Certainly, WAdS$_{3}$ spaces were first obtained as exact solutions to the
equations of motion of TMG \cite{Clement, Clement2}. WAdS$_{3}$ space and
WAdS$_{3}$ black holes are solutions to TMG if the coupling of the
Chern-Simons term,%
\begin{equation}
S_{\text{CS}}=\frac{1}{32\pi G\mu }\int_{\Sigma }d^{3}x\varepsilon ^{\mu \nu
\rho }\Gamma _{\mu \alpha }^{\eta }\left( \partial _{\nu }\Gamma _{\rho \eta
}^{\alpha }+\frac{2}{3}\Gamma _{\nu \beta }^{\alpha }\Gamma _{\rho \eta
}^{\beta }\right) ,
\end{equation}%
and the parameters $\nu $, $l$ satisfy the relation $\nu =\mu l/3$. For WAdS$%
_{3}$ black holes of TMG, it had already been observed in \cite%
{GiribetLeston} that the computation of the mass using the Brown-York tensor
in the boundary yielded the result%
\begin{equation}
M=\dfrac{(\nu ^{2}+3)}{48lG}\left( r_{+}+r_{-}-\dfrac{1}{\nu }\sqrt{(\nu
^{2}+3)r_{+}r_{-}}\right) ,
\end{equation}%
which, again, is in notable agreement with other calculations using
different methods, cf. \cite{Strominger}. In the case of TMG, the large $r$
limit of $Q[\xi ]$ is regularized by introducing a boundary cosmological
term with coefficient 
\begin{equation*}
a_{0}=-\frac{\sqrt{\nu ^{2}+3}}{2l},
\end{equation*}%
which also tends to the AdS$_{3}$ value, $-1/l$, in the limit $\nu =1$.
Therefore, the computation of the WAdS$_{3}$ black hole mass in terms of the
boundary stress-tensor is seen to work in different scenarios.

\subsection{Angular momentum}

Now, let us go back to NMG. In contrast to the computation of the mass $%
M=Q[N_{t}u]$, charge $Q[\partial _{\varphi }]$, which is associated to the
WAdS$_{3}$ black hole angular momentum, does not yield a finite result in
the limit $r\rightarrow \infty $. In fact, boundary terms (\ref{Sc}) do not
suffice to regularize the divergences appearing in the charge $Q[\partial
_{\varphi }]=\int dsu^{i}T_{i\varphi }^{\ast }$. This is simply expressed by
the fact that $u_{\varphi }=0$. Nevertheless, the finite part of the large $%
r $ expansion of $Q[\partial _{\varphi }]$ happens to capture the physically
relevant information. This can be seen by looking at the stress-tensor
expansion%
\begin{equation*}
T_{\varphi \varphi }^{\ast }\simeq t_{\varphi \varphi }^{(1)}r+t_{\varphi
\varphi }^{(0)}+\mathcal{O}(r^{-1}),
\end{equation*}%
which results in an expansion of the form 
\begin{equation*}
Q[\partial _{\varphi }]\simeq J_{(2)}r^{2}+J_{(1)}r+J_{(0)}+\mathcal{O}%
(r^{-1}),
\end{equation*}%
where the coefficient $J_{(2)}$ depends only on $\nu $, while coefficient $%
J_{(1)}$ depends both on $\nu $ and $r_{\pm }$; namely%
\begin{eqnarray*}
J_{(2)} &=&\frac{9}{8}\frac{\nu (\nu ^{2}-1)^{2}}{(20\nu ^{2}-3)l}, \\
J_{(1)} &=&\frac{3}{8}\frac{\nu (\nu ^{2}-1)}{(20\nu ^{2}-3)l}((\nu
^{2}+3)(r_{+}+r_{-})- \\
&&4\nu \sqrt{(\nu ^{2}+3)r_{+}r_{-}}).
\end{eqnarray*}

From this expansion it is not hard to verify that, in contrast to the case
of the mass, for $\nu ^{2}\neq 1$ the introduction of only local
counterterms (\ref{Sc}) does not produce contributions to cancel the
divergences in $Q[\partial _{\varphi }]$. However, remarkably enough, the
finite part $J_{(0)}$ gives the correct result for the black hole angular
momentum; that is, 
\begin{eqnarray}
J_{(0)} &=&\frac{\nu (\nu ^{2}+3)}{4(20\nu ^{2}-3)Gl}((5\nu
^{2}+3)r_{+}r_{-}-  \notag \\
&&2\nu \sqrt{(\nu ^{2}+3)r_{+}r_{-}}(r_{+}+r_{-})).  \label{J}
\end{eqnarray}

To see that (\ref{J}) actually reproduces the correct result one may resort
to the computation done in Ref. \cite{Clement3}, where the
Abbott-Deser-Tekin \cite{ADT, ADT2} method to compute conserved charges was
used to obtain the WAdS$_{3}$ black hole angular momentum. The result
obtained in \cite{Clement3} reads 
\begin{equation}
\tilde{J}=\dfrac{\zeta ^{3}\eta ^{2}}{4Gm^{2}l^{2}}\left( (1-\eta
^{2})\omega ^{2}-\dfrac{\rho _{0}^{2}}{(1-\eta ^{2})}\right) ,
\label{tildeJ}
\end{equation}%
where $\zeta =2\nu $, $\eta =-\sqrt{\nu ^{2}+3}/(2\nu )$, $\omega
=(r_{+}+r_{-}+2\eta \sqrt{r_{+}r_{-}})/(2-2\eta ^{2})$, and $\rho
_{0}^{2}=(r_{+}-r_{-})^{2}/4$. Then, after translating (\ref{tildeJ}) to our
notation, one verifies that (\ref{J}) is proportional to (\ref{tildeJ}),
namely $\tilde{J}=J_{(0)}\zeta ^{2}(1-\eta ^{2})/4$, and the proportionality
factor is precisely the (square of the) one that relates the angular
coordinates $\phi $ used in Ref. \cite{Clement3} and our angular coordinate $%
\varphi $; more precisely, we have $\phi =\varphi \sqrt{\zeta ^{2}(1-\eta
^{2})}/2$. This proportionality factor $\zeta ^{2}(1-\eta ^{2})/4$ is also
explained in Appendix C of Ref. \cite{Tonni}; see equation (C.15) therein.
In conclusion, the finite part of charge $Q[\partial _{\varphi }]$ captures
the physically relevant contribution and gives the correct value of the
black hole angular momentum (\ref{J}). The question remains as to how to
understand the failure in regularizing $Q[\partial _{\varphi }]$ as a
consequence of the abstruse asymptotic structure of WAdS$_{3}$ spaces.

\section{Discussion}

In this paper we have studied holographic renormalization for
three-dimensional massive gravity about WAdS spacetime. The motivation we
had for studying this was to investigate to what extent the standard
holographic renormalization\ techniques can be applied \textit{mutatis
mutandis} to spaces that asymptote WAdS$_{3}$.

The results of our analysis show that the attempt of directly applying the
holographic renormalization recipe to WAdS$_{3}$ spaces partially fails and
partially succeeds:\ While, on one hand, such procedure leads to an exact
computation of conserved charges of asymptotically WAdS$_{3}$ black holes,
it does not suffice to define a fully regularized stress-tensor at the
boundary of the space. Still, it provides a finite definition of the
quasilocal energy density, which gives the right value for the black hole
mass. Also, the precise value of the black hole angular momentum is given by
the finite part of the large radius expansion of the adequate contraction of
the stress-tensor.

At this point, a natural question arises as to why the standard holographic
renormalization technique does not suffice to define a finite boundary
tensor. To this regard, we would like to discuss an interesting possibility:
There is strong evidence that gravity in WAdS$_{3}$ is dual to a
two-dimensional theory that violates Lorentz invariance, being invariant
only under $SL(2,\mathbb{R})\times U(1)$ group. Then, it is natural to ask
whether supplementing $T_{ij}^{\ast }$ with counterterms that do not
preserve Lorentz invariance could ultimately result in a finite boundary
tensor. Precisely because of the lack of Lorentz invariance, such a tensor
would likely be non-symmetric and, in turn, it would not take the form of a
Belinfante tensor associated to improved boundary counterterms as in (\ref%
{Tijren}). Still, the question remains as to whether improving $T_{ij}^{\ast
}$ by adding other kind of Lorentz violating contributions would result in a
regularized boundary quantity. Nevertheless, an exhaustive inspection of all
the contributions one has at hand, which we collect in the Appendix for
completeness, shows that there is no a clear way of improving the boundary
tensor without spoiling the right values of conserved charges.

\begin{equation*}
\end{equation*}

This work has been supported by UBA, CONICET, and ANPCyT. The authors thank
St\'{e}phane Detournay, Rodrigo Olea, and Julio Oliva for interesting
comments, and they specially thank Alan Garbarz and Mauricio Leston for
collaboration and discussions. The authors also thank the referee of JHEP\
for a very interesting suggestion. G.G. is grateful to the members of
Pontificia Universidad Cat\'{o}lica de Valpara\'{\i}so for the hospitality
during his stay;\ specially to Olivera Mi\v{s}kovi\'{c}.

\section{Appendix}

Let us collect the explicit expressions corresponding to the large $r$
expansion of the relevant quantities at the boundary: The components of the
stress tensor, given the expansion $T_{ij}^{\ast
}=t_{ij}^{(1)}r+t_{ij}^{(0)}+t_{ij}^{(-1)}r^{-1}+t_{ij}^{(-2)}r^{-2}+...$,
are given by%
\begin{eqnarray}
t_{tt}^{(0)} &=&\dfrac{\nu ^{2}\sqrt{\nu ^{2}+3}}{\left( 20\nu ^{2}-3\right)
\pi lG},\quad t_{tt}^{(-1)}=0, \\
t_{tt}^{(-2)} &=&\dfrac{\nu ^{2}\sqrt{\nu ^{2}+3}\left( r_{+}-r_{-}\right)
^{2}}{4\left( 20\nu ^{2}-3\right) \pi lG}, \\
t_{t\varphi }^{(1)} &=&\dfrac{3\nu (\nu ^{2}-1)\sqrt{\nu ^{2}+3}}{8(20\nu
^{2}-3)\pi lG}, \\
t_{t\varphi }^{(0)} &=&\dfrac{\nu \sqrt{\nu ^{2}+3}}{16\left( 20\nu
^{2}-3\right) \pi lG}((5\nu ^{2}+3)(r_{+}+r_{-})+  \notag \\
&&8\nu \sqrt{\nu ^{2}+3r_{+}r_{-}}), \\
t_{\varphi \varphi }^{(1)} &=&\dfrac{3\nu (\nu ^{2}-1)\sqrt{\nu ^{2}+3}}{%
8\left( 20\nu ^{2}-3\right) \pi lG}(\nu (r_{+}+r_{-})  \notag \\
&&-\sqrt{\nu ^{2}+3r_{+}r_{-}}), \\
t_{\varphi \varphi }^{(0)} &=&\dfrac{\nu \sqrt{\nu ^{2}+3}}{32(20\nu
^{2}-3)\pi lG}(\nu (13\nu ^{2}+3)(r_{+}^{2}+r_{-}^{2})-  \notag \\
&&2(5\nu ^{2}+3)\sqrt{\nu ^{2}+3r_{+}r_{-}}(r_{+}+r_{-})-  \notag \\
&&2\nu (5\nu ^{2}-21)r_{+}r_{-}).
\end{eqnarray}

On the other hand, the non-vanishing components of boundary metric, denoted
as $\gamma _{ij}\simeq \gamma _{ij}^{(2)}r^{2}+\gamma _{ij}^{(1)}r+\gamma
_{ij}^{(0)}+...$, are the following

\begin{eqnarray}
\gamma _{tt}^{(0)} &=&1, \\
\gamma _{t\varphi }^{(1)} &=&\nu , \\
\gamma _{t\varphi }^{(0)} &=&-\dfrac{1}{2}\sqrt{(\nu ^{2}+3)r_{+}r_{-}}, \\
\gamma _{\varphi \varphi }^{(2)} &=&\dfrac{3}{4}(\nu ^{2}-1), \\
\gamma _{\varphi \varphi }^{(1)} &=&\dfrac{1}{4}(\nu
^{2}+3)(r_{+}+r_{-})-\nu \sqrt{(\nu ^{2}+3)r_{+}r_{-}}.
\end{eqnarray}%
\newline

The components of $\hat{f}_{ij}$, following the same notation, namely $\hat{f%
}_{ij}=\hat{f}_{ij}^{(2)}r^{2}+\hat{f}_{ij}^{(1)}r+\hat{f}_{ij}^{(0)}+...$,
are given by

\begin{eqnarray}
\hat{f}_{tt}^{(0)} &=&-\dfrac{4\nu ^{2}-3}{m^{2}l^{2}}, \\
\hat{f}_{t\varphi }^{(1)} &=&-\dfrac{\nu (4\nu ^{2}-3)}{m^{2}l^{2}}, \\
\hat{f}_{t\varphi }^{(0)} &=&\dfrac{(4\nu ^{2}-3)\sqrt{(\nu ^{2}+3)r_{+}r_{-}%
}}{2m^{2}l^{2}}, \\
\hat{f}_{\varphi \varphi }^{(2)} &=&-\dfrac{9(\nu ^{2}-1)(2\nu ^{2}-1)}{%
4m^{2}l^{2}}, \\
\hat{f}_{\varphi \varphi }^{(1)} &=&\dfrac{(2\nu ^{2}-3)(\nu
^{2}+3)(r_{+}+r_{-})}{4m^{2}l^{2}}+  \notag \\
&&\dfrac{\nu (4\nu ^{2}-3)\sqrt{(\nu ^{2}+3)r_{+}r_{-}}}{m^{2}l^{2}}, \\
\hat{f}_{\varphi \varphi }^{(0)} &=&-\dfrac{3(\nu ^{2}-1)(\nu
^{2}+3)r_{+}r_{-}}{2m^{2}l^{2}}.
\end{eqnarray}%
\newline

Other relevant quantity for the boundary terms is tensor $\nabla _{r}\hat{f}%
_{ij}$, whose components in the large $r$ expansion, $\nabla _{r}\hat{f}%
_{ij}^{(n)}$, are the following

\begin{eqnarray}
\nabla _{r}\hat{f}_{tt}^{(0)} &=&0, \\
\nabla _{r}\hat{f}_{t\varphi }^{(1)} &=&-\dfrac{6\nu (2\nu ^{2}-3)\sqrt{\nu
^{2}+3}r}{m^{2}l^{3}}, \\
\nabla _{r}\hat{f}_{t\varphi }^{(0)} &=&\dfrac{\nu \sqrt{\nu ^{2}+3}(2\nu
^{2}-3)(r_{+}+r_{-})}{2m^{2}l^{3}}, \\
\nabla _{r}\hat{f}_{\varphi \varphi }^{(2)} &=&-\dfrac{2(\nu ^{2}-1)(6\nu
^{2}-9)\sqrt{\nu ^{2}+3}}{4m^{2}l^{3}}, \\
\nabla _{r}\hat{f}_{\varphi \varphi }^{(1)} &=&\dfrac{(2\nu ^{2}-3)\sqrt{\nu
^{2}+3}}{2m^{2}l^{3}}((\nu ^{2}-3)(r_{+}+r_{-})+  \notag \\
&&2\nu \sqrt{(\nu ^{2}+3)r_{+}r_{-}}) \\
\nabla _{r}\hat{f}_{\varphi \varphi }^{(0)} &=&\dfrac{(2\nu ^{2}-3)(\nu
^{2}+3)}{2m^{2}l^{3}}(\sqrt{\nu ^{2}+3}(r_{+}^{2}+r_{-}^{2})-  \notag \\
&&4\nu \sqrt{r_{+}r_{-}}(r_{+}+r_{-})+\sqrt{\nu ^{2}+3}r_{+}r_{-}).
\end{eqnarray}

Contributions to charges $M=2\pi \rho u^{i}T_{ij}^{\ast }\xi ^{j}$ and$\
J=2\pi \rho u^{i}T_{i\varphi }^{\ast }$ coming from the large $r$ expansions 
$M\simeq M_{(1)}r+M_{(0)}+...$ and $J\simeq
J_{(2)}r^{2}+J_{(1)}r+J_{(0)}+... $ are composed as follows: For the mass,
we have

\begin{eqnarray*}
\dfrac{M_{(1)}}{2\pi } &=&\rho ^{(1)}u_{(0)}^{t}t_{tt}^{(0)}+\rho
^{(1)}u_{(0)}^{t}t_{t\varphi }^{(1)}\xi _{-1}^{\varphi }+\rho
^{(1)}u_{(-1)}^{\varphi }t_{\varphi t}^{(1)}=0, \\
\dfrac{M_{(0)}}{2\pi } &=&\rho ^{(1)}u_{(-1)}^{t}t_{tt}^{(0)}+\rho
^{(0)}u_{(0)}^{t}t_{tt}^{(0)}+\rho ^{(1)}u_{(0)}^{t}t_{t\varphi }^{(0)}\xi
_{(-1)}^{\varphi }+ \\
&&\rho ^{(1)}u_{(0)}^{t}t_{t\varphi }^{(1)}\xi _{(-2)}^{\varphi }+\rho
^{(1)}u_{(-1)}^{t}t_{t\varphi }^{(1)}\xi _{(-1)}^{\varphi }+ \\
&&\rho ^{(0)}u_{(0)}^{t}t_{t\varphi }^{(1)}\xi _{(-1)}^{\varphi }+\rho
^{(1)}u_{(-1)}^{\varphi }t_{\varphi t}^{(0)}+\rho ^{(1)}u_{(-2)}^{\varphi
}t_{\varphi t}^{(1)}+ \\
&&\rho ^{(0)}u_{(-1)}^{\varphi }t_{\varphi t}^{(1)}+\rho
^{(1)}u_{(-1)}^{\varphi }t_{\varphi \varphi }^{(1)}\xi _{(-1)}^{\varphi },
\end{eqnarray*}%
where $\rho ^{(n)}$ refer to the components of the large expansion of metric
function (\ref{ro}) in powers of $r$;\ analogously for the large $r$
expansion of $u^{t}=u_{(0)}^{t}+u_{(-1)}^{t}r^{-1}+u_{(-2)}^{t}r^{-2}$, $%
u^{\varphi }\simeq u_{(-1)}^{\varphi }r^{-1}+u_{(-2)}^{\varphi }r^{-2}$, and
the Killing vectors $\xi ^{t}=\xi _{(0)}^{t}=1$, $\xi ^{\varphi }\simeq \xi
_{(-1)}^{\varphi }r^{-1}+\xi _{(-2)}^{\varphi }r^{-2}$. For the angular
momentum, the analogous expression is

\begin{eqnarray*}
\dfrac{J_{(2)}}{2\pi } &=&\rho ^{(1)}u_{(0)}^{t}t_{t\varphi }^{(1)}, \\
\dfrac{J_{(1)}}{2\pi } &=&\rho ^{(1)}u_{(0)}^{t}t_{t\varphi }^{(0)}+\rho
^{(1)}u_{(-1)}^{t}t_{t\varphi }^{(1)}+\rho ^{(0)}u_{(0)}^{t}t_{t\varphi
}^{(1)}+ \\
&&\rho ^{(1)}u_{(-1)}^{\varphi }t_{\varphi t}^{(1)}, \\
\dfrac{J_{(0)}}{2\pi } &=&\rho ^{(1)}u_{(0)}^{t}t_{t\varphi }^{(-1)}+\rho
^{(1)}u_{(-1)}^{t}t_{t\varphi }^{(0)}+\rho ^{(1)}u_{(-2)}^{t}t_{t\varphi
}^{(1)}+ \\
&&\rho ^{(0)}u_{(0)}^{t}t_{t\varphi }^{(0)}+\rho
^{(0)}u_{(-1)}^{t}t_{t\varphi }^{(1)}+\rho ^{(-1)}u_{(0)}^{t}t_{t\varphi
}^{(1)}+ \\
&&+\rho ^{(1)}u_{(-1)}^{\varphi }t_{\varphi \varphi }^{(0)}+\rho
^{(1)}u_{(-2)}^{\varphi }t_{\varphi \varphi }^{(1)}+\rho
^{(0)}u_{(-1)}^{\varphi }t_{\varphi \varphi }^{(1)}.
\end{eqnarray*}%
Then, one finds that the mass of the black hole is given by $M_{(0)}$ while
its angular momentum is given by $J_{(0)}$.

\end{document}